\documentclass[letterpaper, 10 pt, conference]{ieeeconf}  

\IEEEoverridecommandlockouts                              

\usepackage{amsmath,amssymb,amsfonts,mathrsfs}
\usepackage{algorithmic}
\usepackage{algorithm}
\usepackage{array}
\usepackage[caption=false,font=normalsize,labelfont=sf,textfont=sf]{subfig}
\usepackage{textcomp}
\usepackage{stfloats}
\usepackage{url}
\usepackage{verbatim}
\usepackage{graphicx}
\usepackage{cite}

\title{\LARGE \bf
Geometry Enhanced Optimal Control Technique for Acrobatic Flip Motion of Quadcopter
}
\author{Jie Yao 
\thanks{Jie Yao is with Postdoctoral Research Associate of the Department of Mechanical Engineering, University of Minnesota at Twin Cities, MN, USA
        {\tt\small yao00208@umn.edu}}}       
\begin{document}

\maketitle
\thispagestyle{empty}
\pagestyle{empty}

\begin{abstract}
A nonlinear optimal control strategy, named the geometry enhanced finite time $\boldsymbol{\theta-}$D technique, is proposed to manipulate the acrobatic flip flight of variable pitch (VP) quadcopter unmanned aerial vehicles (abbreviated as VP copter). A unique superiority of the VP copter, which can provide the thrust in both positive and negative vertical directions by varying the pitch angles of blades, facilitates the acrobatic flip motion. The finite time $\boldsymbol{\theta-}$D  technique can offer a closed-form near-optimal state feedback control law with online computational efficiency as compared with the finite time state-dependent Riccati equation (SDRE) technique. Meanwhile, by virtue of the geometric technique, the singularity issue of the rotation matrix in the acrobatic flip maneuver can be avoided. The simulation experiments verify the proposed control strategy is effective and efficient. 
\end{abstract}

\section{INTRODUCTION}
Due to its superiority in providing force in both positive and negative directions, VP copter has been receiving wide attention and has been applied to different aspects of our human society \cite{cutler2012design}. Numerous kinds of research focus on the VP copter. Cutler et.al 
\cite{cutler2015analysis} developed a nonlinear, quaternion-based control technique associated with a trajectory planning scheme that can be used to find a polynomial minimum-time trajectory with the constraint of actuator saturation.  Bhargavapuri et.al \cite{bhargavapuri2019robust} presented a backstepping-based nonlinear robust controller to achieve the flip motion of VP copter. Fresh \cite{fresk2013modeling} presented a quaternion-based control method to control the attitude of a VP copter. Sheng and Sun \cite{sheng2016control} proposed an identification method to get the parameters of the dynamic model of the electric variable pitch. In addition, a control strategy and the adaptive compensation method for the VP blades were proposed with consideration the minimize energy consumption. As a result, the response performance of the lift force was increased, and the direct lift-based flight control method also extremely improved flight performance. \\
\indent Manipulating the acrobatic motion of VP copter i.e. flip flight is attractive but challenging. Cutler and How \cite{cutler2012actuator} presented a trajectory generation method by fitting polynomials (including the waypoints) in 3-dimensional space. Attitude-oriented constraints were incorporated into the polynomial trajectory formulation, leading to aerobatic maneuvers to be conducted by using a single controller and trajectory generation algorithm. Gupta et.al. \cite{gupta2017modeling} employed a control architecture with three loops, i.e. outer loop, inner loop, and control allocation loop, to achieve the flip motion.  Gan et.al \cite{yu2019quadrotor} presented a control method to perform the aggressive flight (360° flip) with satisfying attitude-constrained conditions. The first step was to generate the trajectory by a quadratic programming problem with some linear constraints. After that, a tracking controller was developed to follow the generated trajectory. Meanwhile, the asymptotic stability of the closed-loop tracking system can be guaranteed.  \\
\indent Undoubtedly, it is more challenging to perform the agile flight based on the perspectives of "optimality", "finite time", and "singularity-free". In this manuscript, a control architect, which incorporates attributes of the finite time, optimality, and singularity-free, is proposed and is named as geometry-enhanced finite time $\boldsymbol{\theta-}$D scheme. Actually, developing a closed-form state-feedback optimal control law for a nonlinear system is difficult since there exists a technical barrier that the partial differential Hamilton-Jacobi-Bellman (HJB) can not be solved analytically \cite{lewis2012optimal}. Admittedly, some numerical methods can generate numerical optimal solutions to nonlinear systems. Rao published a comprehensive review paper \cite{rao2009survey} about the numerical optimal control methodologies. The key limitation of the numerical methods is that only open-loop solutions can be accessed. The state-dependent Riccati equation technique is a beautiful scheme, which can provide a closed-form state-feedback optimal control law. However, it suffers intensive online computational burdens. Based on those considerations, the proposed finite time $\boldsymbol{\theta-}$D scheme can get a closed-form state-feedback optimal control law with less online computational load, significantly facilitating the onboard implementation. Lee \cite{lee2010geometric} introduced the geometric concept into UAV attitude control. Since the geometric concept is a coordinate-free control strategy, the singularities and complexities of using local coordinates can be circumvented. \\
\indent The contribution of this manuscript is summarized as follows:
Both advantages of the computational efficiency of finite time $\boldsymbol{\theta-}$D technique and singularity-free property in the geometrical method are fully incorporated and exploited to develop a new control architecture, which can manipulate the VP copter to fulfill acrobatic flip and drive it from start point to end point successfully. Compared with the finite-time SDRE technique, the simulation results show that the proposed control scheme is effective. More significantly, it can save almost 50 \% of computing time which is necessary for the corresponding SDRE control method.\\
\indent The remainder of the paper is organized as follows: Section II presents the dynamics of VP copter. Section III introduces the control allocation mechanism. The main procedures of finite time $\boldsymbol{\theta-}$D technique are illustrated in Section IV. Section V describes the geometry-enhanced finite time $\boldsymbol{\theta-}$D scheme-based controller design. Section VI is the numerical results and analysis. The concluding remark is shown in Section VII.
\section{VARIABLE PITCH  QUADCOPTER DYNAMIC MODEL}
In this manuscript, the gesture of VP copter is described over an inertial coordinate O-XYZ. At the center of mass of (CoM) the copter, a body coordinate o-xyz is attached to it. The translational profile is expressed within the inertial frame by:\\
\noindent \textbf{Position}:
\begin{equation}\label{1}
    \boldsymbol{\mathcal{X}} = [x_c, y_c, z_c]^T
\end{equation}
\textbf{Linear Velocity}:
\begin{equation}\label{2}
    \Dot{\boldsymbol{\mathcal{X}}} = [\Dot{x}_c, \Dot{y}_c, \Dot{z}_c]^T
\end{equation}
\indent The attitude profile can be described by:\\
\textbf{Orientation Angle}:
\begin{equation}\label{3}
    \boldsymbol{\Omega} = [\Phi, \Theta, \Psi]^T
\end{equation}
\textbf{Angular Rate}:
\begin{equation}\label{4}
    \Dot{\boldsymbol{\Omega}} = [\Dot{\Phi}, \Dot{\Theta}, \Dot{\Psi}]^T
\end{equation}
\indent The \textbf{thrust force} $\boldsymbol{T}_{vp}$ is defined into the body frame and along with the z-axis. The \textbf{torque}, $\boldsymbol{\tau}_{vp} = [ \tau_{\Phi}, \tau_{\Theta}, \tau_{\Psi} ]$, is depicted on the inertial frame along with angle $[\Phi, \Theta, \Psi]$. \\
\indent The overall state $\boldsymbol{x}$ will be collected as
\begin{align}\label{5}
    \boldsymbol{x}  & = \Big[ \boldsymbol{\mathcal{X}}^T,  \Dot{\boldsymbol{\mathcal{X}}}^T, \boldsymbol{\Omega}^T,  \Dot{\boldsymbol{\Omega}}^T \Big]^T \nonumber \\
    & = \Big[x_c, y_c, z_c, \Dot{x}_c, \Dot{y}_c, \Dot{z}_c, \Phi, \Theta, \Psi,  \Dot{\Phi}, \Dot{\Theta}, \Dot{\Psi}  \Big]^T_{12 \times 1}
\end{align}
\indent Then, the dynamics of the VP copter can be modeled as a state-space form as \cite{nekoo2019optimized}, 
\begin{align}\label{6}
    \Dot{\boldsymbol{x}} & = \left [
    \begin{array}{c}
         \Dot{\boldsymbol{\mathcal{X}}}  \\
         \Ddot{\boldsymbol{\mathcal{X}}}  \\
         \Dot{\boldsymbol{\Omega}} \\
         \Ddot{\boldsymbol{\Omega}} 
    \end{array} \right]  \\
    & = \left[
    \begin{array}{cc}
         \Dot{\boldsymbol{\mathcal{X}}}  \\
         \frac{1}{m_{vp}}\boldsymbol{\textit{I}}_{3 \times 3} \bigg( R_{vp}(:,3) \boldsymbol{T}_{vp} - m_{vp}g\boldsymbol{e}_3 - D_{vp} \Dot{\boldsymbol{\mathcal{X}}}    \bigg) \\
         \Dot{\boldsymbol{\Omega}} \\
         J_{vp}^{-1}(\boldsymbol{\Omega})\bigg( \boldsymbol{\tau}_{vp} - C_{vp}(\boldsymbol{\Omega}, \Dot{\boldsymbol{\Omega}}) \Dot{\boldsymbol{\Omega}} \bigg)
    \end{array}\right] \nonumber
\end{align}
Some parameters in Eq.(\ref{6}) are explained as follows:\\
1). $m_{vp}$ is the mass of copter;\\
2). $\boldsymbol{\textit{I}}_{3 \times 3}$ is $3 \times 3$ identity matrix;\\
3). $g$ is the gravity acceleration constant; \\
4). $\boldsymbol{e}_3$ is denoted as $\boldsymbol{e}_3 = [0,0,1]^T$; \\
5). $D_{vp}$ is the aerodynamic effect \cite{luukkonen2011modelling} and represents as $D_{vp} = \textit{diag} \big( [D_x, D_y, D_z] \big)$; \\
6). $R_{vp}(:,3)$ is the third column of rotation matrix $R_{vp}(t)$, and it is defined in the special orthogonal group $SO(3)$. The property of  $SO(3)$ is 
\begin{align}\label{7}
    &SO(3) \triangleq   \nonumber \\
    &\Big\{ R_{vp}(t) \in \mathbb{R}^{3 \times 3} |R_{vp}^TR_{vp} =  \boldsymbol{\textit{I}}_{3 \times 3}, det[R_{vp}(t)] = 1 \Big\}
\end{align}
\indent In addition, the variation of $R_{vp}(t)$ is 
\begin{equation}\label{8}
    \Dot{R}_{vp}(t) = R_{vp}(t) \widehat{\Pi}
\end{equation}
where $\Pi$ represents the angular velocity, and the hat operator is $\widehat{\Pi} = [\Pi\times]$,which means the skew symmetric matrix. \\
\indent It is worth noting that the rotation matrix $R_{vp}(t)$ from $SO(3)$ benefits from the philosophy of geometric control described in \cite{lee2010geometric}. With the kinematic equation of Eq.(\ref{8}), the singularity issue can be circumvented in the control design process.\\
7). The inertial matrix $J_{vp}$ and Coriolis term $C_{vp}(\boldsymbol{\Omega}, \Dot{\boldsymbol{\Omega}})$ can be found at the \cite{nekoo2019optimized}. The detailed descriptions are skipped here since they are well-developed concepts.  \\
\indent According to Eq.(6), the thrust force $\boldsymbol{T}_{vp}$ and torque $\boldsymbol{\tau}_{vp}$ are regarded as the control commands to VP copter dynamics of Eq.(\ref{6}). In real scenarios, the servo motors will receive commands from ESC (electronic speed controller) to change the angles of the propellers. Thus, thrust force $\boldsymbol{T}_{vp}$ and torque $\boldsymbol{\tau}_{vp}$ will be varied according to the principle of aerodynamics. Usually, the first step is to develop the relationship between blade angles to thrust coefficients based on the blade element theory and momentum theory. The second step is to utilize some motion mechanisms to describe the mapping of the thrust coefficient to the thrust force $\boldsymbol{T}_{vp}$ \& torque $\boldsymbol{\tau}_{vp}$.\\
\indent Actually, the relationship of blade angles to thrust coefficients is well-developed \cite{seddon2011basic}, and its expression is, 
\begin{equation}\label{9}
    \alpha_i(t) = \frac{3}{2} \gamma_{vp} + \bigg(\frac{6}{\sigma_{vp} C_{l\alpha}^{vp}}\bigg) C_{t_i}^{vp}(t)
\end{equation}
\indent The parameters of Eq.(\ref{9}) are described as follows:\\
1). The $i^{\text{th}}$ blade angle is $\alpha_i$ $(i = 1, 2, 3, 4)$; \\
2). $C_{t_i}^{vp}$ represents the thrust coefficient from the $i^{\text{th}}$ blade;\\
3). $C_{l\alpha}^{vp}$ denotes the airfoil lift curve slope; \\
4). $\sigma_{vp}$ is calculated by the equation of 
\begin{equation}\label{10}
    \sigma_{vp} = N_b^{vp} \frac{c_{vp}}{\pi r_{vp}}
\end{equation}
where $N_b^{vp}$ is the number of blades, $r_{vp}$ is the tip radius of the blade, and $c_{vp}$ is the chord length;\\
5). $\gamma_{vp}$ is the inflow ratio.\\
\indent The next step is to define the relationship between the thrust coefficients and the thrust force $\boldsymbol{T}_{vp}$ \& torque $\boldsymbol{\tau}_{vp}$. Based on the effects of the blade angle variation on the thrust and torques \cite{bhargavapuri2019robust}, their relationship can be defined as 
\begin{small}
\begin{align}\label{11}
&\left[
\begin{array}{c}
     \boldsymbol{T}_{vp}\\
     \tau_{\Phi}\\
     \tau_{\Theta}\\
     \tau_{\Psi}
\end{array}\right]=  \\
&\left[
\begin{array}{cccc}
     K_{vp} & K_{vp} & K_{vp} & K_{vp} \\
     0 & -L_{vp}K_{vp} & 0 & L_{vp}K_{vp} \\
     -L_{vp}K_{vp} & 0 & L_{vp}K_{vp} & 0 \\
     -\tilde{a}\sqrt{\vert C_{t_1}^{vp}\vert} & \tilde{a}\sqrt{\vert C_{t_2}^{vp}\vert} & -\tilde{a}\sqrt{\vert C_{t_3}^{vp}\vert} & \tilde{a}\sqrt{\vert C_{t_4}^{vp}\vert }
\end{array}
\right]
\underbrace{
\left[
\begin{array}{c}
     C_{t_1}^{vp}\\
     C_{t_2}^{vp}\\
     C_{t_3}^{vp}\\
     C_{t_4}^{vp}
\end{array}\right]}_{\boldsymbol{C}_{vp}} \nonumber
\end{align}
\end{small}

\noindent where $K_{vp} \triangleq \rho_{air}\pi r_{vp}^4 \omega_{ss}^2$, in which $\rho_{air}$ is the air density and $\omega_{ss}$ is the steady-state angular velocity of the blade, $\tilde{a} \triangleq (r_{vp}K_{vp})/\sqrt{2}$, and $L_{vp}$ is the length of the rotor to the CoM.

\section{CONTROL ALLOCATION MECHANISM}
According to Eq.(9) and Eq.(11), it is obvious that there exists a \textbf{nonlinear mapping} from blade angles to thrust \& torque. Then, a proper dynamic control allocation strategy is necessary, i.e.  pseudo-inverse method, augmented pseudo-inverse method, first order dynamics, mean value theorem, and null-space of pseudo-inverse approach.\cite{bhargavapuri2019robust,nekoo2019optimized}. In this paper, we employ the augmented pseudo-inverse method. Firstly, Eq.(11) needs to be separated linear part of Eq.(\ref{12}) and the nonlinear part of Eq.(\ref{13}). 
\begin{align}\label{12}
    &  \underbrace{\left[
    \begin{array}{c}
     \boldsymbol{T}_{vp}\\
     \tau_{\Phi}\\
     \tau_{\Theta}\\
    \end{array}
    \right]}_{\boldsymbol{H}_1} =  \\
& \underbrace{\left[
\begin{array}{cccc}
     K_{vp} & K_{vp} & K_{vp} & K_{vp} \\
     0 & -L_{vp}K_{vp} & 0 & L_{vp}K_{vp} \\
     -L_{vp}K_{vp} & 0 & L_{vp}K_{vp} & 0 
\end{array}
\right]}_{\boldsymbol{M}_1}
\underbrace{
\left[
\begin{array}{c}
     C_{t_1}^{vp}\\
     C_{t_2}^{vp}\\
     C_{t_3}^{vp}\\
     C_{t_4}^{vp}
\end{array}\right]}_{\boldsymbol{C}_{vp}} \nonumber
\end{align}

\begin{align}\label{13}
        \tau_{\Psi}/\tilde{a} = \underbrace{\bigg[-\sqrt{\vert C_{t_1}^{vp}\vert}, \sqrt{\vert C_{t_2}^{vp}\vert}, -\sqrt{\vert C_{t_3}^{vp}\vert}, \sqrt{\vert C_{t_4}^{vp}\vert} \bigg]\boldsymbol{C}_{vp}}_{\kappa(\boldsymbol{C}_{vp})}
\end{align}
\indent Then, an optimization problem is formulated as 
\begin{align}\label{14}
    \underbrace{\text{min}}_{\boldsymbol{C}_{vp}} \frac{1}{2} \boldsymbol{C}_{vp}^T W_{vp} \boldsymbol{C}_{vp}
\end{align}
with the constraints of Eq.(\ref{12}) of $\big[ \boldsymbol{H}_1 - \boldsymbol{M}_1 \boldsymbol{C}_{vp} = 0\big] $ and Eq.(\ref{13}) of  $\big[ \tau_{\Psi}/\tilde{a} - \kappa(\boldsymbol{C}_{vp}) = 0 \big] $. The $W_{vp} > 0$ is the weight matrix. It should be noted that minimizing the $\boldsymbol{C}_{vp}$ is to indirectly minimize the control effect based on Eq.(\ref{11}). \\
\indent To solve this nonlinear optimization problem, an augment Lagrangian function is organized as 
\begin{align}\label{15}
    \mathcal{L}_{vp} &= \frac{1}{2} \bigg[ \boldsymbol{C}_{vp}^T W_{vp} \boldsymbol{C}_{vp} + \mu_{\Psi} \Big( \tau_{\Psi}/\tilde{a} - \kappa(\boldsymbol{C}_{vp}) \Big)^2   \bigg] \nonumber \\
    &+ \boldsymbol{\Lambda}_{vp}^T\Big( \boldsymbol{H}_1 - \boldsymbol{M}_1 \boldsymbol{C}_{vp}\Big)
\end{align}
where $\boldsymbol{\Lambda}_{vp}$ is the Lagrangian multiplier vector, and $\mu_{\Psi} > 0$ is a scalar value that can determine the nonlinearity's effect.\\
\indent Taking the derivative w.r.t $\boldsymbol{C}_{vp}$ on both sides of Eq.(\ref{15}) and equating it to zero, it leads to
\begin{align}\label{16}
    W_{vp} \boldsymbol{C}_{vp} - \mu_{\Psi} \Big( \tau_{\Psi}/\tilde{a} - \kappa(\boldsymbol{C}_{vp}) \Big) \frac{\partial \kappa(\boldsymbol{C}_{vp})}{\partial \boldsymbol{C}_{vp}} - \boldsymbol{M}_1^T\boldsymbol{\Lambda}_{vp} = 0
\end{align}
where 
\begin{align*}
 &\frac{\partial \kappa(\boldsymbol{C}_{vp})}{\partial \boldsymbol{C}_{vp}}  = \widetilde{\boldsymbol{f}}(\boldsymbol{C}_{vp})\boldsymbol{C}_{vp},\hspace{5pt} \text{and} \hspace{5pt} \widetilde{\boldsymbol{f}}(\boldsymbol{C}_{vp}) = \nonumber \\
 &\frac{3}{2} \text{diag}\Bigg( -\frac{\text{sign}(C_{t_1}^{vp})}{\sqrt{\vert C_{t_1}^{vp}\vert}}, \frac{\text{sign}(C_{t_2}^{vp})}{\sqrt{\vert C_{t_2}^{vp}\vert}}, 
 -\frac{\text{sign}(C_{t_3}^{vp})}{\sqrt{\vert C_{t_3}^{vp}\vert}}, \frac{\text{sign}(C_{t_4}^{vp})}{\sqrt{\vert C_{t_4}^{vp}\vert}}\Bigg)
\end{align*}
\indent Then, $\boldsymbol{C}_{vp}$ can be derived from Eq.(\ref{16}), 
\begin{equation}\label{17}
    \boldsymbol{C}_{vp} = \Bigg[\underbrace{W_{vp} - \mu_{\Psi} \Big( \tau_{\Psi}/\tilde{a} - \kappa(\boldsymbol{C}_{vp}) \Big) \widetilde{\boldsymbol{f}}(\boldsymbol{C}_{vp})}_{\Upsilon(\boldsymbol{C}_{vp})} \Bigg]^{-1} \boldsymbol{M}_1^T \boldsymbol{\Lambda}_{vp}
\end{equation}
\indent Taking Eq.(\ref{17}) into Eq.(\ref{12}), one can get
\begin{equation}\label{18}
    \boldsymbol{H}_1 = \boldsymbol{M}_1 \Big[\Upsilon(\boldsymbol{C}_{vp})\Big]^{-1} \boldsymbol{M}_1^T \boldsymbol{\Lambda}_{vp}
\end{equation}
\indent The Lagrangian multiplier $\boldsymbol{\Lambda}_{vp}$ can be obtained from Eq.(\ref{18}) to be 
\begin{equation}\label{19}
    \boldsymbol{\Lambda}_{vp} = \Big( \boldsymbol{M}_1 [\Upsilon(\boldsymbol{C}_{vp})]^{-1} \boldsymbol{M}_1^T \Big)^{-1} \boldsymbol{H}_1
\end{equation}
\indent Finally, one can get $\boldsymbol{C}_{vp}$ by substituting Eq.(\ref{19}) into Eq.(\ref{17}), 
\begin{align}\label{20}
    &\boldsymbol{C}_{vp}(t)_{4 \times 1} = \nonumber \\ &\Big[\Upsilon(\boldsymbol{C}_{vp})\Big]^{-1}  \boldsymbol{M}_1^T \Big( \boldsymbol{M}_1 [\Upsilon(\boldsymbol{C}_{vp})]^{-1} \boldsymbol{M}_1^T \Big)^{-1} \boldsymbol{H}_1
\end{align}
\indent Then, substituting each element Eq.(\ref{20}) into Eq.(\ref{9}), the each blade angle $\alpha_i$ can be calculated. However, it has a physical limitation for blade angle variations, i.e. $\alpha_{max} < \alpha_i < \alpha_{min}$, where $\alpha_{max}$ and $\alpha_{min}$ are an upper bound and lower bound of the blade angle variation. If the calculated $\alpha_i$ exceeds the bounds, $\alpha_{max}$ and $\alpha_{min}$ need to be used to replace the calculated $\alpha_i$ and to be involved in the ensuing calculations. After that, using this $\alpha_i$ to compute the corresponding $C_{t_i}^{vp}$ as
\begin{equation}\label{21}
    C_{t_i}^{vp}(t) = \bigg(\frac{\sigma_{vp} C_{l\alpha}^{vp}}{6}\bigg) \bigg( \alpha_i(t) - \frac{3}{2} \gamma_{vp}\bigg) 
\end{equation}
\indent Thus, this $ C_{t_i}^{vp}(t) $ can be used as input command within Eq.(\ref{11}) to calculate the corresponding thrust and torque. 
\section{THE DEVELOPMENT OF FINITE TIME $\boldsymbol{\theta-}$D SUBOPTIMAL CONTROL}
\indent Considering a class of nonlinear dynamics as 
\begin{equation}\label{22}
    \Dot{\boldsymbol{x}} = \boldsymbol{f(x)} + \boldsymbol{g(x)} \boldsymbol{u}
\end{equation}
with the cost function as 
\begin{align}\label{23}
    \mathrm{J} = \frac{1}{2} \boldsymbol{x}^T(t_f)\widetilde{S}\boldsymbol{x}(t_f) + \frac{1}{2}\int_{0}^{t_f} \bigg( \boldsymbol{x}^T\widetilde{Q}\boldsymbol{x} + \boldsymbol{u}^T \widetilde{R} \boldsymbol{u} \bigg) dt 
\end{align}
where $\boldsymbol{x}$, $\boldsymbol{f}$, $\boldsymbol{g}$, and $\boldsymbol{u}$ are evolving within appropriate dimensional compact sets which are subsets of Euclidean space. The penalty matrices $\widetilde{S}$, $\widetilde{Q}$ and $\widetilde{R}$ have compatible dimensions with $\widetilde{S} \geq 0$, $\widetilde{Q} \geq 0$ and $\widetilde{R} > 0$. \\
\indent The optimal control law is 
\begin{equation}\label{24}
    \boldsymbol{u} = -\widetilde{R}^{-1}\boldsymbol{g}^T(\boldsymbol{x})\mathbb{V}_{\boldsymbol{x}}
\end{equation}
where $\mathbb{V}_{\boldsymbol{x}}$ needs to be acquired from the following Hamilton-Jacobi-Bellman (HJB) equation, 
\begin{equation}\label{25}
    \mathbb{V}^T_{\boldsymbol{x}}\boldsymbol{f(x)} - \frac{1}{2}\mathbb{V}^T_{\boldsymbol{x}}\boldsymbol{g(x)}\widetilde{R}^{-1}\boldsymbol{g}^T(\boldsymbol{x})\mathbb{V}_{\boldsymbol{x}} + \frac{1}{2}\boldsymbol{x}^T\widetilde{Q}\boldsymbol{x} = - \mathbb{V}_t
\end{equation}
where $\mathbb{V}_{\boldsymbol{x}} = \partial \mathbb{V}(\boldsymbol{x},t)/ \partial \boldsymbol{x}$, $\mathbb{V}_t = \partial \mathbb{V}(\boldsymbol{x},t)/ \partial t$, and $\mathbb{V}(\boldsymbol{x},t)$ is the optimal cost-to-go, i.e.
\begin{align}\label{26}
    &\mathbb{V}(\boldsymbol{x},t)  \\
    & = \underbrace{\text{min}}_{\boldsymbol{u}} \bigg\{ \frac{1}{2} \boldsymbol{x}^T(t_f)\widetilde{S}\boldsymbol{x}(t_f) + \frac{1}{2}\int_{t}^{t_f} \bigg( \boldsymbol{x}^T\widetilde{Q}\boldsymbol{x} + \boldsymbol{u}^T \widetilde{R} \boldsymbol{u} \bigg) dt \bigg\} \nonumber 
\end{align}
\indent It is worth noting that the analytical solutions of Eq.(\ref{25}) can not be solved if the dynamics are nonlinear, which restricts the application scope of optimal control methodologies to numerous nonlinear systems. Actually, SDRE is an attractive and potential scheme to approximate the solutions of the HJB equation. However, it has an unavoidable defect, i.e. intensive online computational load. Interested readers can refer to the paper \cite{heydari2013path}, in which one can find that at every time instant, an algebraic Riccati equation, matrix exponential, and twice matrix inverse is required to be calculated. The proposed $\boldsymbol{\theta-}$D scheme is to be developed to compensate for this deficiency. \\
\indent The state penalty matrix $\widetilde{Q}$ in Eq.(\ref{23}) should be modified as $\{ \widetilde{Q} + \sum_{i=1}^{\infty}\boldsymbol{\theta}^i D_i \}$, in which the $\boldsymbol{\theta} > 0$ is a supplemental scalar variable, and $\{ \widetilde{Q} + \sum_{i=1}^{\infty}\boldsymbol{\theta}^i D_i \}$ should be positive semidefinite with proper selection of $D_i$ terms.\\
\indent Eq.(\ref{22}) should be rewritten to be 
\begin{align}\label{27}
    \Dot{\boldsymbol{x}}= \underbrace{\Big[\boldsymbol{A}_0  + \boldsymbol{\theta} \bigg( \frac{\boldsymbol{A(x)}}{\boldsymbol{\theta}}\bigg)\Big]\boldsymbol{x}}_{\boldsymbol{f(x)}}
    + \underbrace{\Big[\boldsymbol{B}_0  + \boldsymbol{\theta} \bigg( \frac{\boldsymbol{B(x)}}{\boldsymbol{\theta}}\bigg)\Big]}_{\boldsymbol{g(x)}}\boldsymbol{u}
\end{align}
where $\Big( \boldsymbol{A}_0, \boldsymbol{B}_0  \Big)$ are constant matrix pair, and stabilizable; $\Big( \boldsymbol{A}_0 + \boldsymbol{A(x)}, \boldsymbol{B}_0 + \boldsymbol{B(x)} \Big)$ is point-wise controllable \cite{ccimen2010systematic}. \\
\indent The $\mathbb{V}_{\boldsymbol{x}}$ is assumed to be expended as 
\begin{equation}\label{28}
    \mathbb{V}_{\boldsymbol{x}} = \bigg( \mathbb{T}_0(t) + \sum_{i=1}^{\infty} \mathbb{T}_i(\boldsymbol{x}) \boldsymbol{\theta}^i \bigg) \boldsymbol{x}
\end{equation}
where $\{\mathbb{T}_i, i = 0, \cdots, \infty\}$ are assumed to be symmetric and can be determined at the following derivations. \\
\indent Based on Eq.(\ref{28}), the optimal cost-to-go $\mathbb{V}(\boldsymbol{x},t)$ can be obtained from $\mathbb{V}_{\boldsymbol{x}}$ by integral operation,  
\begin{align}\label{29}
    \mathbb{V} & = \int  \mathbb{V}_{\boldsymbol{x}} ^T d\boldsymbol{x} + \mathcal{C}'(t) \nonumber \\
    & = \int  \Bigg[ \bigg( \mathbb{T}_0(t) + \sum_{i=1}^{\infty} \mathbb{T}_i(\boldsymbol{x}) \boldsymbol{\theta}^i \bigg) \boldsymbol{x} \Bigg] ^T d\boldsymbol{x} + \mathcal{C}'(t) \nonumber \\
    & = \frac{1}{2}\boldsymbol{x}^T\mathbb{T}_0(t)\boldsymbol{x} + \mathcal{C}''(\boldsymbol{x}, \boldsymbol{\theta}) + \mathcal{C}'(t) 
\end{align}    
where $\mathcal{C}'(t)$ is function of $t$ and determined by the boundary condition, and $\mathcal{C}''(\boldsymbol{x}, \boldsymbol{\theta}) = \int \Big[ \big( \sum_{i=1}^{\infty} \mathbb{T}_i(\boldsymbol{x}) \boldsymbol{\theta}^i \big) \boldsymbol{x} \Big] ^T d\boldsymbol{x} $ is function of $\boldsymbol{x}$ and $ \boldsymbol{\theta}$. \\
\indent In addition, two assumptions, $\{ \mathcal{C}'(t) = 0 | t \in [0,t_f] \} $ and $\{ \mathcal{C}''(\boldsymbol{x}, \boldsymbol{\theta}) = 0 | t = t_f \} $, should be hold. Admittedly, those two assumptions are strong and at the expense of optimal solutions to Eq.(\ref{22}) and Eq.(\ref{23}). However, it is similar to the SDRE design scheme \cite{ccimen2010systematic} within which $\mathbf{P}(\boldsymbol{x},t)\boldsymbol{x}$ is assumed to be the gradient of the optimal cost-to-go $\mathbb{V}(\boldsymbol{x},t)$ and $\mathbf{P}(\boldsymbol{x},t)$ is also assumed to be symmetric, which may sacrifice the optimality. Due to those assumptions, the proposed finite time $\boldsymbol{\theta-}$D is claimed to be a sub-optimal or near-optimal control design strategy. \\
\indent Since we have known the expression of Eq.(\ref{29}), then the $\mathbb{V}_t$ can be derived as 
\begin{equation}\label{30}
    \mathbb{V}_t = \frac{\partial \mathbb{V}}{\partial t} = \frac{1}{2}\boldsymbol{x}^T\Dot{\mathbb{T}}_0(t)\boldsymbol{x}
\end{equation}
\indent Additionally, with Eq.(\ref{26}), Eq.(\ref{29}) and two assumptions, we can define the value of $\mathbb{T}(t)$ at the final time $t=t_f$ as 
\begin{equation}\label{31}
    \mathbb{T}_0(t_f) = \widetilde{S}
\end{equation}
\indent Finally, \textbf{substituting} Eq.(\ref{28}), Eq.(\ref{30}), and the decomposition forms of $\boldsymbol{f(x)}$ \& $\boldsymbol{g(x)}$ of Eq.(27) into the Eq.(25),  \textbf{replacing} $\widetilde{Q}$ of Eq.(\ref{25}) with $\{ \widetilde{Q} + \sum_{i=1}^{\infty}\boldsymbol{\theta}^i D_i \}$, \textbf{sorting} all terms based on the power of $\boldsymbol{\theta}$, \textbf{equating} their coefficients to be zeros, it will lead to 
\begin{equation}\label{32}
    \mathbb{T}_0(t)\boldsymbol{A}_0 + \boldsymbol{A}^T_0 \mathbb{T}_0(t) -  \mathbb{T}_0(t) \boldsymbol{B}_0 \widetilde{R}^{-1} \boldsymbol{B}_0^T \mathbb{T}_0(t) + \widetilde{Q} = - \Dot{\mathbb{T}}_0(t)
\end{equation}
\begin{align}\label{33}
    &\mathbb{T}_1(\boldsymbol{A}_0 - \boldsymbol{B}_0\widetilde{R}^{-1}\boldsymbol{B}_0^T\mathbb{T}_0) + (\boldsymbol{A}_0^T - \mathbb{T}_0\boldsymbol{B}_0\widetilde{R}^{-1}\boldsymbol{B}_0^T)\mathbb{T}_1 \nonumber \\
    & = -\frac{\mathbb{T}_0\boldsymbol{A}(\boldsymbol{x})}{
    \boldsymbol{\theta}}-\frac{\boldsymbol{A}^T(\boldsymbol{x})\mathbb{T}_0}{\boldsymbol{\theta}} + \mathbb{T}_0 \boldsymbol{B}_0 \widetilde{R}^{-1} \frac{\boldsymbol{B}^T(\boldsymbol{x})}{\boldsymbol{\theta}}\mathbb{T}_0 \nonumber \\
    &+ \mathbb{T}_0\frac{\boldsymbol{B}(\boldsymbol{x})}{\boldsymbol{\theta}}\widetilde{R}^{-1}\boldsymbol{B}_0^T\mathbb{T}_0 - D_1
\end{align}
\begin{align}\label{34}
    &\mathbb{T}_2(\boldsymbol{A}_0 - \boldsymbol{B}_0\widetilde{R}^{-1}\boldsymbol{B}_0^T\mathbb{T}_0) + (\boldsymbol{A}_0^T - \mathbb{T}_0\boldsymbol{B}_0\widetilde{R}^{-1}\boldsymbol{B}_0^T)\mathbb{T}_2 \nonumber \\   
    & = -\frac{\mathbb{T}_1\boldsymbol{A}(\boldsymbol{x})}{
    \boldsymbol{\theta}}-\frac{\boldsymbol{A}^T(\boldsymbol{x})\mathbb{T}_1}{\boldsymbol{\theta}} + \mathbb{T}_0 \boldsymbol{B}_0 \widetilde{R}^{-1} \frac{\boldsymbol{B}(\boldsymbol{x})^T}{\boldsymbol{\theta}}\mathbb{T}_1 \nonumber \\
    & + \mathbb{T}_0\frac{\boldsymbol{B}(\boldsymbol{x})}{\boldsymbol{\theta}}\widetilde{R}^{-1}\boldsymbol{B}_0^T\mathbb{T}_1 
    + \mathbb{T}_0\frac{\boldsymbol{B}(\boldsymbol{x})}{\boldsymbol{\theta}}\widetilde{R}^{-1}\frac{\boldsymbol{B}^T(\boldsymbol{x})}{\boldsymbol{\theta}}\mathbb{T}_0 \nonumber \\
    & + \mathbb{T}_1\boldsymbol{B}_0\widetilde{R}^{-1}\boldsymbol{B}_0^T\mathbb{T}_1 
    + \mathbb{T}_1\boldsymbol{B}_0\widetilde{R}^{-1}\frac{\boldsymbol{B}^T(\boldsymbol{x})}{\boldsymbol{\theta}}\mathbb{T}_0 \nonumber\\
    & + \mathbb{T}_1\frac{\boldsymbol{B}(\boldsymbol{x})}{\boldsymbol{\theta}}\widetilde{R}^{-1}\boldsymbol{B}_0^T\mathbb{T}_0 - D_2,
\end{align}
\begin{align*}
   \vdots   
\end{align*}
\indent The general formula can be organized in a compact form as, 
\begin{align}\label{35}
        &\mathbb{T}_i(\boldsymbol{A}_0 - \boldsymbol{B}_0\widetilde{R}^{-1}\boldsymbol{B}_0^T\mathbb{T}_0) + (\boldsymbol{A}_0^T - \mathbb{T}_0\boldsymbol{B}_0\widetilde{R}^{-1}\boldsymbol{B}_0^T)\mathbb{T}_i \nonumber \\   
        & = -\frac{\mathbb{T}_{i-1}\boldsymbol{A}(\boldsymbol{x})}{
    \boldsymbol{\theta}}-\frac{\boldsymbol{A}^T(\boldsymbol{x})\mathbb{T}_{i-1}}{\boldsymbol{\theta}}  \nonumber \\
        & + \sum_{j=0}^{i-1} \mathbb{T}_j\Bigg( \boldsymbol{B}_0\widetilde{R}^{-1}\frac{\boldsymbol{B}^T(\boldsymbol{x})}{\boldsymbol{\theta}} + \frac{\boldsymbol{B}(\boldsymbol{x})}{\boldsymbol{\theta}} \widetilde{R}^{-1}\boldsymbol{B}^T_0\Bigg)\mathbb{T}_{i-1-j}\nonumber \\ 
    & + \sum_{j=0}^{i-2} \mathbb{T}_j\frac{\boldsymbol{B}(\boldsymbol{x})}{\boldsymbol{\theta}}\widetilde{R}^{-1}\frac{\boldsymbol{B}^T(\boldsymbol{x})}{\boldsymbol{\theta}}\mathbb{T}_{i-2-j}\nonumber \\
    & + \sum_{j=1}^{i-1} \mathbb{T}_j\boldsymbol{B}_0\widetilde{R}^{-1}\boldsymbol{B}^T_0\mathbb{T}_{i-j}- D_i.    
\end{align}
\indent  The $D_i$ terms is designed as 
\begin{equation}\label{36}
    D_i = p_i e^{-q_it} \Big\{ \text{Right-Hand-Side of Eq.(\ref{35})}\Big\}
\end{equation}
where the index $i$ can pick as $\{ i = 1, \cdots,\infty\}$.\\
\indent Substituting Eq.(\ref{36}) into Eq.(\ref{35}), it yields the general equation of $\mathbb{T}_i$ with perturbed $D_i$ terms, 
\begin{align}\label{37}
    &\mathbb{T}_i(\boldsymbol{A}_0 - \boldsymbol{B}_0\widetilde{R}^{-1}\boldsymbol{B}_0^T\mathbb{T}_0) + (\boldsymbol{A}_0^T - \mathbb{T}_0\boldsymbol{B}_0\widetilde{R}^{-1}\boldsymbol{B}_0^T)\mathbb{T}_i \nonumber \\   
        & = \underbrace{\bigg( 1 - p_i e^{-q_it}\bigg)}_{\{\rho_i(t)|i = 1, \cdots,\infty\} }\Bigg\{-\frac{\mathbb{T}_{i-1}\boldsymbol{A}(\boldsymbol{x})}{
    \boldsymbol{\theta}}-\frac{\boldsymbol{A}^T(\boldsymbol{x})\mathbb{T}_{i-1}}{\boldsymbol{\theta}}  \nonumber \\
        & + \sum_{j=0}^{i-1} \mathbb{T}_j\Bigg( \boldsymbol{B}_0\widetilde{R}^{-1}\frac{\boldsymbol{B}^T(\boldsymbol{x})}{\boldsymbol{\theta}} + \frac{\boldsymbol{B}(\boldsymbol{x})}{\boldsymbol{\theta}} \widetilde{R}^{-1}\boldsymbol{B}^T_0\Bigg)\mathbb{T}_{i-1-j}\nonumber \\ 
    & + \sum_{j=0}^{i-2} \mathbb{T}_j\frac{\boldsymbol{B}(\boldsymbol{x})}{\boldsymbol{\theta}}\widetilde{R}^{-1}\frac{\boldsymbol{B}^T(\boldsymbol{x})}{\boldsymbol{\theta}}\mathbb{T}_{i-2-j}\nonumber \\
    & + \sum_{j=1}^{i-1} \mathbb{T}_j\boldsymbol{B}_0\widetilde{R}^{-1}\boldsymbol{B}^T_0\mathbb{T}_{i-j}\Bigg\}   
\end{align}
\indent Finally, the original control law of Eq.(\ref{24}) can be updated as 
\begin{equation}\label{38}
    \boldsymbol{u}_{\theta-\text{D}} = -\widetilde{R}^{-1}\boldsymbol{g}^T(\boldsymbol{x}) \bigg( \mathbb{T}_0(t) + \sum_{i=1}^{\infty} \mathbb{T}_i(\boldsymbol{x}) \boldsymbol{\theta}^i \bigg) \boldsymbol{x} 
\end{equation}
\indent Some superiority features should be emphasized as follows: Eq.(\ref{32}) can be solved offline since $\boldsymbol{A}_0$, $\boldsymbol{B}_0$, $\widetilde{R}$, $\widetilde{Q}$ and $\mathbb{T}_0(t_f)$ are known constant matrices,  leading to the solutions $\mathbb{T}_0(t)$ can be solved with backward integration, and then be stored for the following calculations. One can symbolize the right hand side of Eq.(\ref{37}) with $\widetilde{W}(\boldsymbol{x},\theta,p_i, q_i, t)$. Then, according to the matrix analysis theory, the solutions $\mathbb{T}_i$ of Eq.(\ref{37}) (which is a typical \textbf{linear Lyapunov equation}) can be expressed as 
\begin{align}\label{39}
   & Vec(\mathbb{T}_i)) =
   \bigg( I_n \otimes (\boldsymbol{A}_0^T - \mathbb{T}_0\boldsymbol{B}_0\widetilde{R}^{-1}\boldsymbol{B}_0^T)  \nonumber \\
  & + 
   (\boldsymbol{A}_0^T - \mathbb{T}_0\boldsymbol{B}_0\widetilde{R}^{-1}\boldsymbol{B}_0^T) \otimes I_n \bigg) \widetilde{W}(\boldsymbol{x},\theta,p_i,q_i,t)
\end{align}
where the symbol of $Vec$ is the vectorization operator \cite{golub2013matrix}, and the term $\big( I_n \otimes (\boldsymbol{A}_0^T - \mathbb{T}_0\boldsymbol{B}_0\widetilde{R}^{-1}\boldsymbol{B}_0^T)  
  + (\boldsymbol{A}_0^T - \mathbb{T}_0\boldsymbol{B}_0\widetilde{R}^{-1}\boldsymbol{B}_0^T) \otimes I_n \big)$
can be calculated offline. It is worth emphasizing that only matrix addition and multiplication operations are necessary to be performed online. It is the theoretical foundation for why the proposed $\boldsymbol{\theta-}$D can significantly reduce the online computational time as compared with the corresponding SDRE scheme. In addition, the perturbation term $D_i$ (or $\rho_i(t)$ term in Eq.(\ref{37})) is used to avoid the large control gain caused by the large initial values. Furthermore, in real implementations, the existence of the $D_i$ term can protect the actuators from overloading. Finally, for most engineering problems, only three terms $\mathbb{T}_0(t)$, $\mathbb{T}_1(\boldsymbol{x},\boldsymbol{\theta})$, and $\mathbb{T}_2(\boldsymbol{x},\boldsymbol{\theta})$ are sufficient to guarantee the satisfactory performances \cite{yao2022finite}.
\section{GEOMETRY ENHANCED FINITE TIME $\boldsymbol{\theta-}$D BASED CONTROL DESIGN FOR VP COPTER}
Geometry-enhanced finite time $\boldsymbol{\theta-}$D-based controller will be developed to force the VP copter 
from the initial position to reach a target position at the prescribed final time. Meanwhile, the flip maneuver will be operated. The control design process is decomposed into \textbf{translational control design} and \textbf{attitude control design}. \\
\indent The first and second equations of Eq.(\ref{6}) represents the translational dynamics of VP copter. A new state $\boldsymbol{x}_t$ is defined as $\Dot{\boldsymbol{x}}_t = [\boldsymbol{\mathcal{X}}, \Dot{\boldsymbol{\mathcal{X}}} ]^T$ with a state-space equation as 
 \begin{align}\label{40}
    \dot{\boldsymbol{x}}_t & = 
    \underbrace{
    \left[
    \begin{array}{cc}
        0_{3 \times 3} & I_{3 \times 3} \\
        0_{3 \times 3} & - (1/m_{vp})I_{3 \times 3}D_{vp}
    \end{array}
    \right]}_{\mathbf{A}_t}
    \underbrace{
        \left[
    \begin{array}{c}
        \boldsymbol{\mathcal{X}} \\
        \Dot{\boldsymbol{\mathcal{X}}}
    \end{array}
    \right]}_{\mathbf{x}_t} \nonumber \\
    & +
    \underbrace{
     \left[
    \begin{array}{c}
        0_{3 \times 3} \\
        I_{3 \times 3}
    \end{array}
    \right]}_{\boldsymbol{B}_t}
    \underbrace{\Big[(1/m_{vp})R_{vp}(:,3)\boldsymbol{T}_{vp} - g\boldsymbol{e}_3\Big]}_
    {\boldsymbol{u}_t} 
\end{align}
\indent Then, the translational control law can be 
\begin{equation}\label{41}
    \boldsymbol{u}_t = -\widetilde{R}_t^{-1}\boldsymbol{B}_t^T\boldsymbol{K}_t(\boldsymbol{x}_t - \boldsymbol{x}_{t,des})
\end{equation}
where $\boldsymbol{x}_{t,des}$ is desired position and speed. By selecting proper $\widetilde{R}_t$, $\widetilde{Q}_t$ and $\widetilde{S}_t$, $\boldsymbol{K}_t$ can be calculated by only solving the Eq.(\ref{32})  since $\mathbf{A}_t$ and $\boldsymbol{B}_t$ are constant matrices. After getting the $\boldsymbol{u}_t$, $\boldsymbol{T}_{vp}$ has to been recalculated with
\begin{align}\label{42}
    \boldsymbol{T}_{vp} = m_{vp} \Big ( R_{vp}(1,3) \boldsymbol{u}_t(1) + R_{vp}(2,3) \boldsymbol{u}_t(2) \nonumber \\
    + R_{vp}(3,3) (\boldsymbol{u}_t(3) + g)\Big)
\end{align}
\indent The third and fourth equations of Eq.(\ref{6}) represents the attitude dynamics of VP copter. A new state $\boldsymbol{x}_t$ is defined as $\Dot{\boldsymbol{x}}_a = [\boldsymbol{\Omega}, \Dot{\boldsymbol{\Omega}} ]^T$ with a state-space equation as 
\begin{align}\label{43}
    \dot{\boldsymbol{x}}_a & = 
    \underbrace{
    \left[
    \begin{array}{cc}
        0_{3 \times 3} & I_{3 \times 3} \\
        0_{3 \times 3} & -J_{vp}^{-1}(\boldsymbol{x}_a)C_{vp}(\boldsymbol{x}_a)
    \end{array}
    \right]}_{\boldsymbol{A}_a(\boldsymbol{x}_a)}
    \underbrace{
        \left[
    \begin{array}{c}
        \boldsymbol{\Omega} \\
        \Dot{\boldsymbol{\Omega}}
    \end{array}
    \right]}_{\boldsymbol{x}_a} \nonumber \\
    & +
    \underbrace{
     \left[
    \begin{array}{c}
        0_{3 \times 3} \\
        J_{vp}^{-1}(\boldsymbol{x}_a)
    \end{array}
    \right]}_{\boldsymbol{B}_a(\boldsymbol{x}_a)}
    \underbrace{[\boldsymbol{\tau}_{vp}]}_{\boldsymbol{u}_a} 
\end{align}
\indent Based on Eq.(43), the geometry-enhanced finite time $\boldsymbol{\theta-}$D based attitude controller can be designed as 
\begin{equation}\label{44}
    \boldsymbol{u}_a = -\widetilde{R}_a^{-1}\boldsymbol{B}^T_a(\boldsymbol{x}_a)
    \Big(\mathbb{T}_0 + \theta \mathbb{T}_1(\boldsymbol{x}_a) + \theta^2 \mathbb{T}_1(\boldsymbol{x}_a)\Big)
    \left[
    \begin{array}{c}
         \boldsymbol{e}_R  \\
         \boldsymbol{e}_{\Omega}
    \end{array}
    \right]
\end{equation}
\indent By selecting proper $\widetilde{R}_a$, $\widetilde{Q}_a$ and $\widetilde{S}_a$ and using the Eq.(\ref{32}) to Eq.(\ref{34}),  $\boldsymbol{u}_a$ can be obtained in which \textbf{the error signal $\boldsymbol{[\boldsymbol{e}_R, \boldsymbol{e}_{\Omega}]^T}$ are from the geometric technique \cite{lee2017geometric}\big(distinct from the traditional error signal $(\boldsymbol{x}_a - \boldsymbol{x}_{a,des})$\big)}.\\
\indent According to the geometric technique, an error function over the nonlinear space SO(3) \cite{lee2010geometric} should be defined,
\begin{equation}\label{45}
    \boldsymbol{E}(R_{vp}(t), R_{d}(t)) = 2-\sqrt{\text{trace}\Big( R_d^T(t) R_{vp}(t) \Big)+1}
\end{equation}
where $\text{trace}(\bullet)$ is trace operator, and $R_d^T(t)$ is the desired rotation matrix.\\
\indent The variation of Eq.(\ref{45}) is
\begin{align}\label{46}
    \Delta_{R_{vp}(t)}\Big( R_{vp}(t), R_{d}(t) \Big)= \frac{ \partial \boldsymbol{E}(R_{vp}(t), R_{d}(t))}{\partial R_{vp}(t)} \Dot{R}_{vp}(t)
\end{align}
\indent With the dynamics of $R_{vp}(t)$ Eq.(\ref{8}), Eq.(\ref{46}) leads to 
\begin{align}\label{47}
    \Delta_{R_{vp}(t)}\Big( R_{vp}(t), R_{d}(t) \Big) = \frac{\text{trace}\Big( R_d^T(t) R_{vp}(t)\widehat{\Pi}(t)\Big)}{2\sqrt{\text{trace}\Big( R_d^T(t) R_{vp}(t) \Big)+1}}
\end{align}
\indent The following property can be applied for the numerator of Eq.(\ref{47})
\begin{align}\label{48}
    &\text{trace}\Big( R_d^T(t) R_{vp}(t)\widehat{\Pi}(t)\Big) \nonumber \\
    &= \Big( R_d^T(t) R_{vp}(t) - R_{vp}^T(t) R_{d}(t)\Big)^{\vee} \widehat{\Pi}(t)
\end{align}
where $\vee$ is vee operator \cite{lee2010geometric}. \\
\indent Eq.(\ref{47}) leads to 
\begin{align}\label{48}
    & \Delta_{R_{vp}(t)}\Big( R_{vp}(t), R_{d}(t) \Big)  = \nonumber \\
    & \frac{\Big( R_d^T(t) R_{vp}(t) - R_{vp}^T(t) R_{d}(t)\Big)^{\vee} }{2\sqrt{\text{trace}\Big( R_d^T(t) R_{vp}(t) \Big)+1}} \widehat{\Pi}(t)
\end{align}
\indent Thus, the attitude error $\boldsymbol{e}_R (t)$ can be expressed as 
\begin{equation}\label{49}
    \boldsymbol{e}_R (t) = \frac{\Big( R_d^T(t) R_{vp}(t) - R_{vp}^T(t) R_{d}(t)\Big)^{\vee} }{2\sqrt{\text{trace}\Big( R_d^T(t) R_{vp}(t) \Big)+1}} 
\end{equation}
\indent Besides, $\boldsymbol{e}_{\Omega}$ can be obtained by the following equation,  
\begin{align}\label{50}
    &\Dot{R}_{vp}(t) - \Dot{R}_{d}^T(t)\Big(R_d^T(t) R_{vp}(t) \Big) \nonumber \\
    &=R_{vp}(t) \Big( \Pi(t) - R_{vp}^T(t) R_d(t)\Pi_d(t) \Big)^{\wedge}
\end{align}
\indent Thus, $\boldsymbol{e}_{\Omega}$ is 
\begin{equation}\label{51}
    \boldsymbol{e}_{\Omega}(t) = \Pi(t) - R_{vp}^T(t) R_d(t)\Pi_d(t)
\end{equation}

\indent In this paper, the desired rotation matrix is given by,
\begin{equation}\label{53}
    R_d(t) = [R_d(:,1)(t), R_d(:,2)(t), R_d(:,3)(t)]
\end{equation}
where,
\begin{equation}\label{54}
    R_d(:,1)(t)=
    \left[
    \begin{array}{c}
         c_{\Theta_{des}}c_{\Psi_{des}}  \\
         c_{\Theta_{des}}s_{\Psi_{des}}\\
         -s_{\Theta_{des}}
    \end{array}
    \right]
\end{equation}

\begin{equation}\label{55}
    R_d(:,2)(t)=
    \left[
    \begin{array}{c}
         -s_{\Psi_{des}}c_{\Phi_{des}} + s_{\Theta_{des}}s_{\Phi_{des}}c_{\Psi_{des}}  \\
           s_{\Phi_{des}}c_{\Psi_{des}} s_{\Theta_{des}}+c_{\Psi_{des}}c_{\Phi_{des}}\\
        s_{\Phi_{des}} c_{\Theta_{des}}
    \end{array}
    \right]
\end{equation}

\begin{equation}\label{56}
    R_d(:,3)(t)=
    \left[
    \begin{array}{c}
           s_{\Theta_{des}}c_{\Phi_{des}}c_{\Psi_{des}} + s_{\Psi_{des}}s_{\Phi_{des}}  \\
         -c_{\Psi_{des}}s_{\Phi_{des}} + c_{\Phi_{des}}s_{\Psi_{des}}s_{\Theta_{des}}\\
         c_{\Phi_{des}}c_{\Theta_{des}}
    \end{array}
    \right]
\end{equation}
in which $\Phi_{des}$ and $\Theta_{des}$ can be found in Eq.(\ref{57}) and Eq.(\ref{58}) and $\Psi_{des}$ can be given independently; the symbol $s_{(\bullet)}$ is the sinusoidal function of $sin(\bullet)$, the symbol $c_{(\bullet)}$ indicates cosine function of $cos(\bullet)$.
\begin{equation}\label{57}
    \Theta_{des}(t) = arctan \bigg( \frac{ \boldsymbol{u}_t(2)s_{\Psi_{des}} +\boldsymbol{u}_t(1)c_{\Psi_{des}} }{\boldsymbol{u}_t(3) + g} \bigg)
\end{equation}

\begin{equation}\label{58}
    \Phi_{des}(t) = arcsin \bigg( \frac{ \boldsymbol{u}_t(1)s_{\Psi_{des}} - \boldsymbol{u}_t(2)c_{\Psi_{des}} }{\boldsymbol{u}_t(1)^2 + \boldsymbol{u}_t(2)^2 + \big(\mathbf{u}_t(3) + g\big)^2} \bigg)
\end{equation}
where $\boldsymbol{u}_t(i)$ can be found in Eq.(\ref{41}).\\
\indent To conduct the flip motion within $[0, t_f]$, a flip time slot is selected as $t_{flip} = [t_1, t_2]$, $0 < t_1 < t_2 < t_f$. In this manuscript, it is assumed that the flip maneuver is about the X-axis with the following conditions,
\begin{align}\label{59}
    & \Phi_{des}(t)  = \Phi_{des}(t) ,\hspace{7pt} \text{if} \hspace{5pt} 0 \leq t < t_1; \nonumber \\
    & \Phi_{des}(t)  = \phi_f ,\hspace{7pt} \text{if} \hspace{5pt} t_1 \leq t < t_2;  \\
    & \Phi_{des}(t)  = \Phi_{des}(t) + \phi_f , \hspace{7pt} \text{if} \hspace{5pt} t_2 \leq t \leq t_f; \nonumber 
\end{align}
where $\phi_f$ is the flip value which is set as $\pi$, implying a full flip.
\section{NUMERICAL EXPERIMENT AND ANALYSIS}
A Macbook Pro laptop with a processor i7 and 16 GB memory is used to perform the simulations. There are two objectives of the simulation experiment: 1). verifying the proposed geometry-enhanced $\boldsymbol{\theta}-$D control strategy is effective to manipulate the VP copter to have a full flip along the trajectory to the target position. 2). showing the proposed method is more efficient than the SDRE technique while the performances generated by both techniques are comparable. \\
\indent The VP copter's physical parameters can refer to \cite{rafee2022geometric}. The penalty matrices are selected as $\widetilde{R}_t = I_{3 \times 3}$, $\widetilde{Q}_t = diag([1 \times I_{3 \times 3}, 0 \times I_{3 \times 3}])$, $\widetilde{S}_t = diag([10\times I_{3 \times 3}, 0 \times I_{3 \times 3}])$, $\widetilde{R}_a = I_{3 \times 3}$, $\widetilde{Q}_a = diag([10\times I_{3 \times 3}, 5 \times I_{3 \times 3}])$ and $\widetilde{S}_a = diag([100\times I_{3 \times 3}, I_{3 \times 3}])$. The parameters for $\boldsymbol{\theta}-D$ method are $p_1 = 0.9$, $p_2 = 0.99$, $q_1 = 10$, and $q_2 = 100$. The initial value is $\boldsymbol{x}(0) = 0_{12 \times 1}$. The final condition is $\boldsymbol{x}(t_f) = [-3,2,1, \Phi_{des}(t_f), \Theta_{des}(t_f), 0, 0_{1\times 6}]^T_{12 \times 1}$. \\
\begin{figure}[htbp]
\centering
\includegraphics[scale = 0.055]{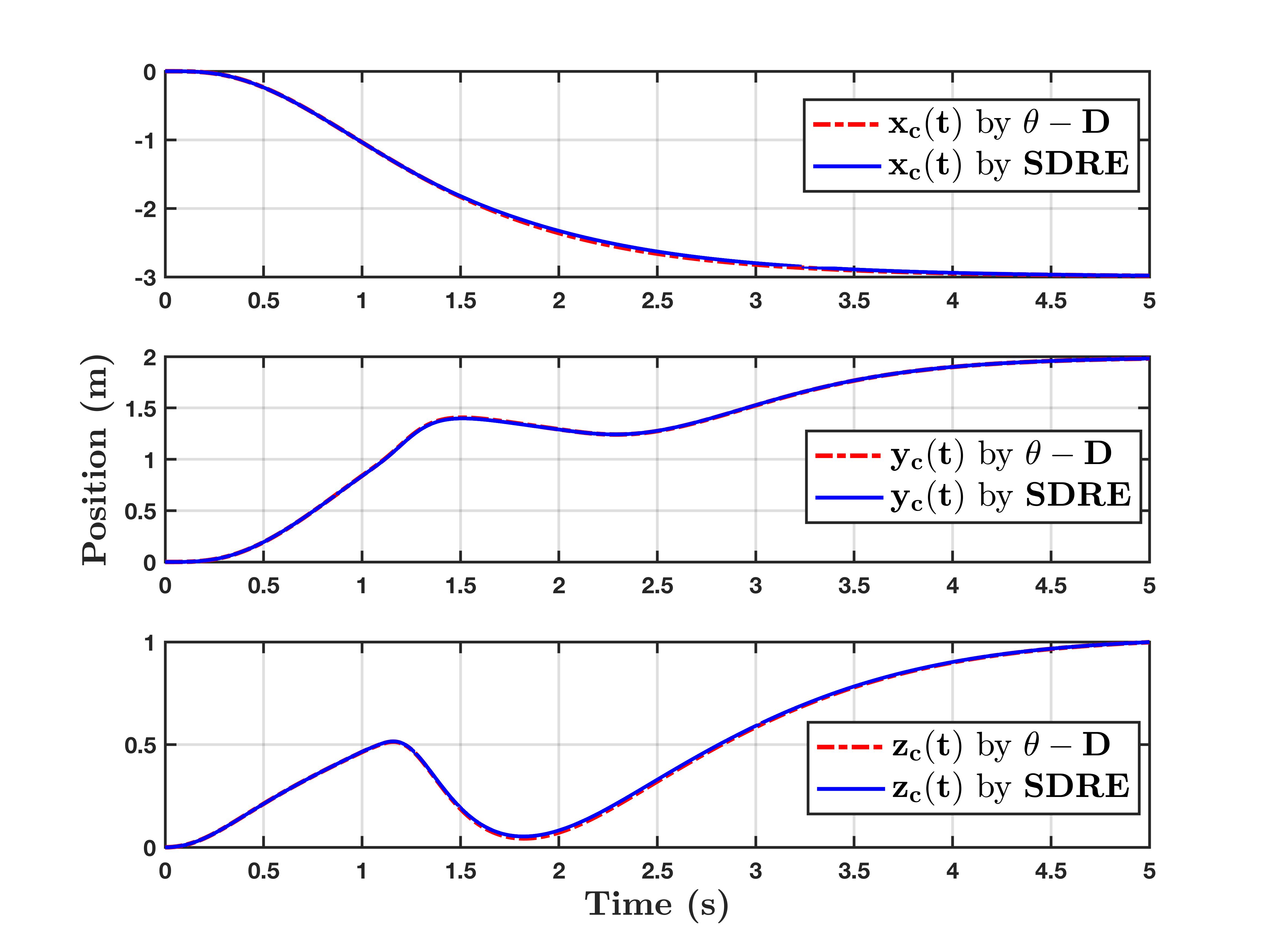}
\caption{ Position profile by finite-time $\boldsymbol{\theta}-$D  \& finite-time SDRE techniques}
\label{Fig.1}
\end{figure}
\indent Figs.\ref{Fig.1}-\ref{Fig.4} shows the performance profile of both methods, including the history of the position, attitude, linear velocity, and angular velocity, in which the red dashed line represents the performance of the proposed method, and the blue line is the performance of the SDRE technique. From those figures, one can observe that both performances are comparable, which implies both techniques are sufficient to drive the VP copter from the initial position to the ending position. Besides, one can notice the performance of $\Phi$, which is varied from zero rad to $\pi$ rad, i.e. a full flip completed. Furthermore, it is also worth pointing out that the singularity issue does not happen due to the property of the geometric control strategy.\\
\indent Fig.\ref{Fig.5} and Fig.\ref{Fig.6} are the thrust and torque profiles generated by both control methods. When comparing the control energy $\int_{0}^{t_f}(\boldsymbol{u}^T\boldsymbol{u})dt$ of both methods, the SDRE consumes 545 units to complete the task while the proposed method requires 544.2 units. Significantly, \textbf{the proposed method only needs 9e-4 seconds to calculate the control commands at every time instant while the SDRE technique requires 1.8e-3 seconds to get the control law  at every time instant.} It clearly demonstrates that the proposed finite-time $\boldsymbol{\theta}-$D technique is more efficient for online implementation. \\
\indent Fig.\ref{7} visualizes the VP copter trajectory produced by the proposed finite time $\boldsymbol{\theta}-$D technique in a three-dimensional format. It is obvious to notice that the positions of the blue rotor and green rotor are totally changed between the start point and end point. The flip process has been highlighted. \\
\section{CONCLUSION}
This manuscript presents a control strategy that combines the merit of the finite time $\boldsymbol{\theta}-$D technique, which is more efficient in online computation,  and the merit of the geometric method, which can circumvent the singularity issue of traditional attitude control for UAV. Numerical simulations verify the proposed control strategy is effective and efficient. Meanwhile, the finite time $\boldsymbol{\theta}-$D strategy can extend to other engineering problems which require fast control executions. 
\section*{ACKNOWLEDGMENT}
The author would like to deeply appreciate Dr. S. N. Balakrishnan's advice in preparing this manuscript.

\begin{figure}
\centering
\includegraphics[scale = 0.055]{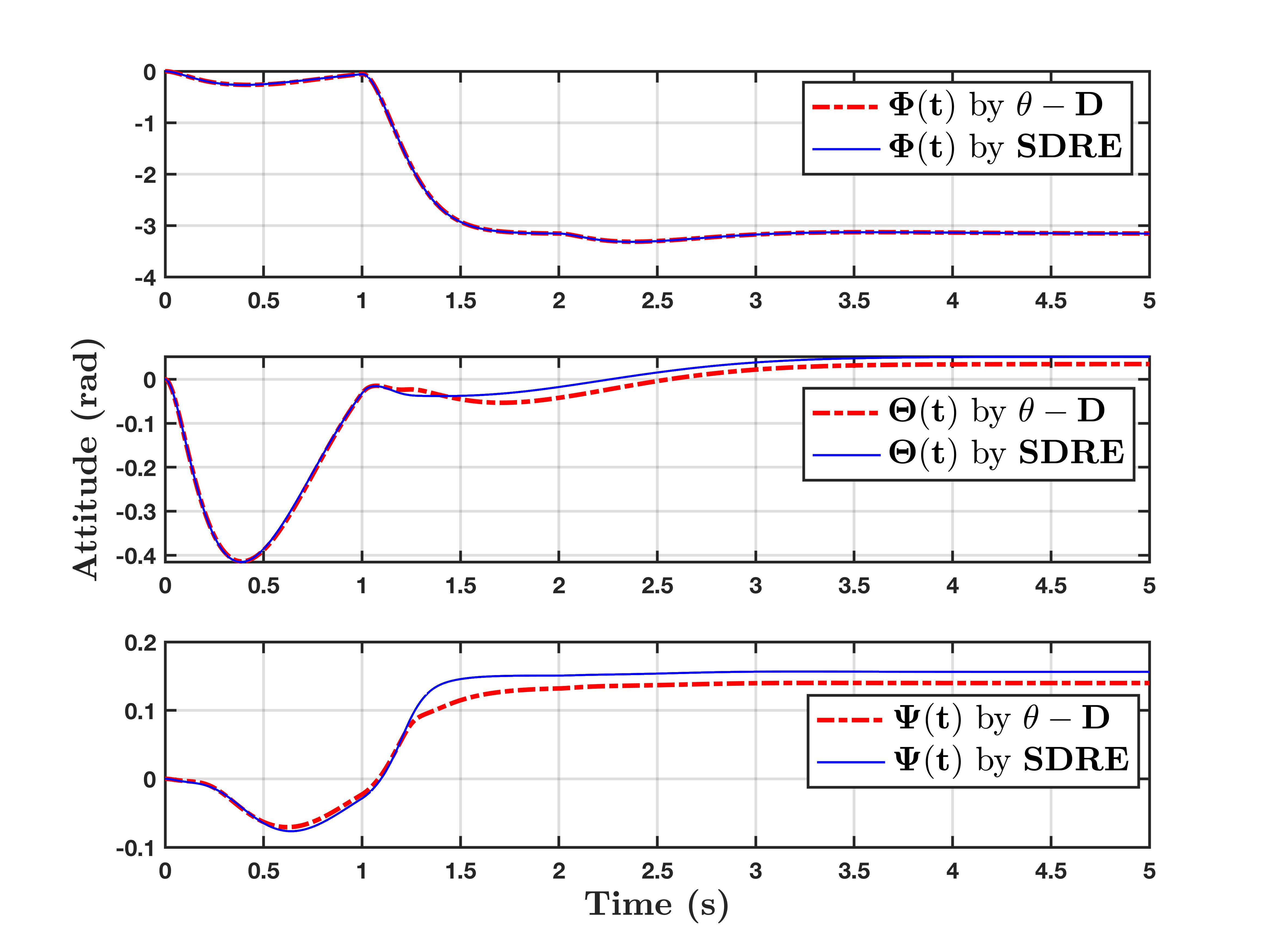}
\caption{ Attitude profile by two techniques}
\label{Fig.2}
\end{figure}
\begin{figure}
\centering
\includegraphics[scale = 0.055]{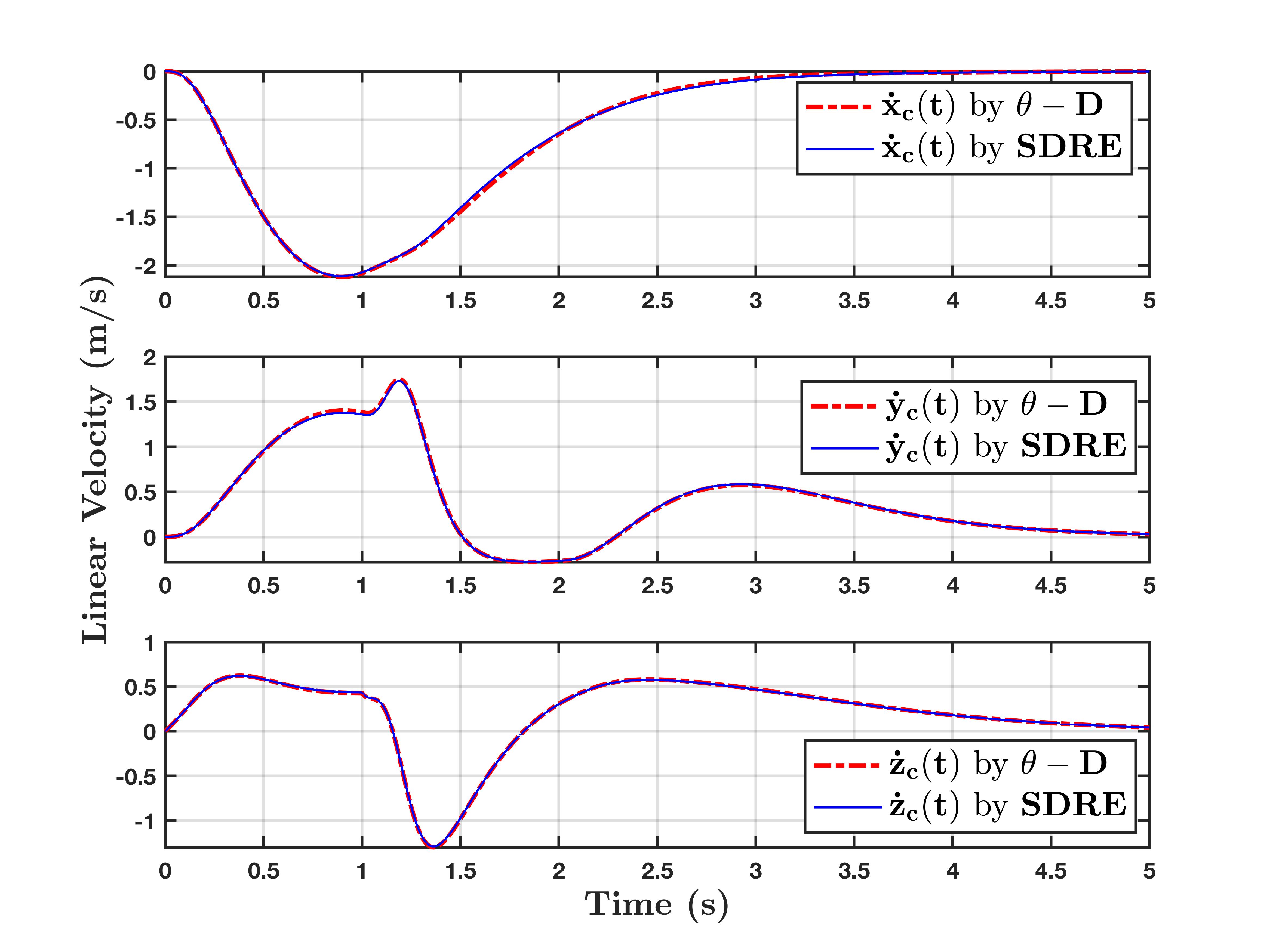}
\caption{ Velocity profile by two techniques}
\label{Fig.3}
\end{figure}
\begin{figure}
\centering
\includegraphics[scale = 0.055]{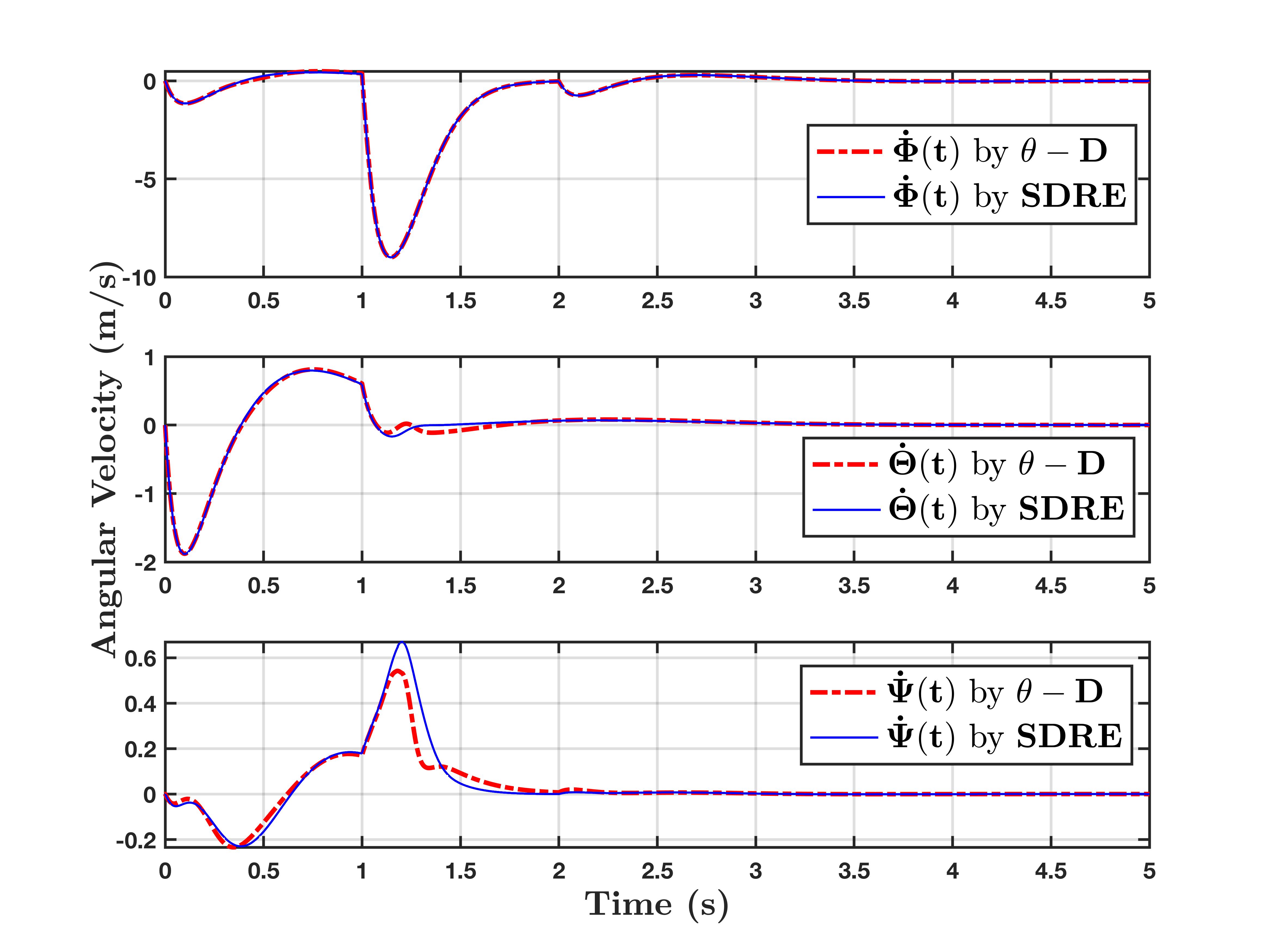}
\caption{ Angular velocity profile by two techniques}
\label{Fig.4}
\end{figure}
\begin{figure}
\centering
\includegraphics[scale = 0.055]{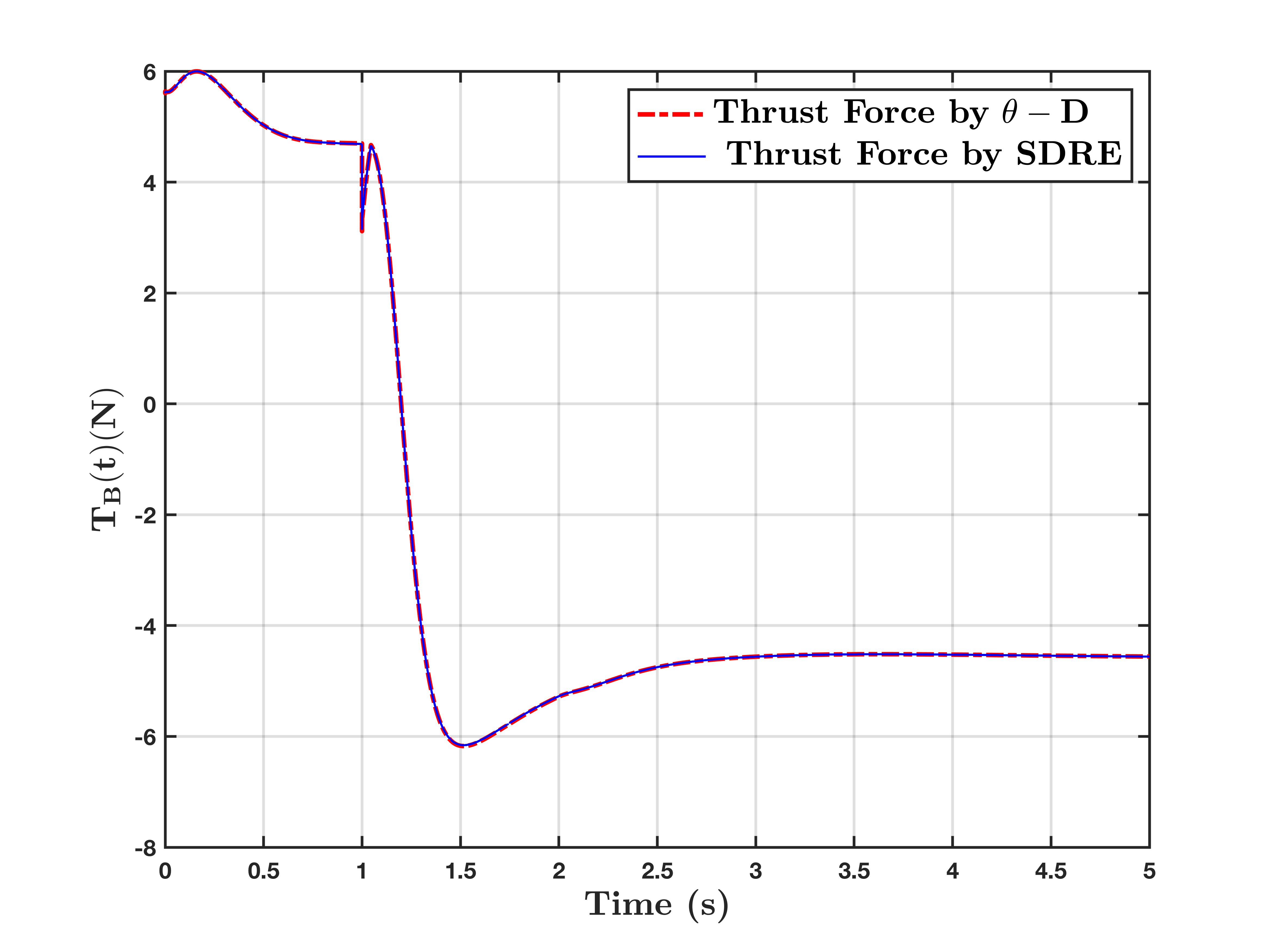}
\caption{ Thrust force profile by two techniques}
\label{Fig.5}
\end{figure}
\begin{figure}
\centering
\includegraphics[scale = 0.055]{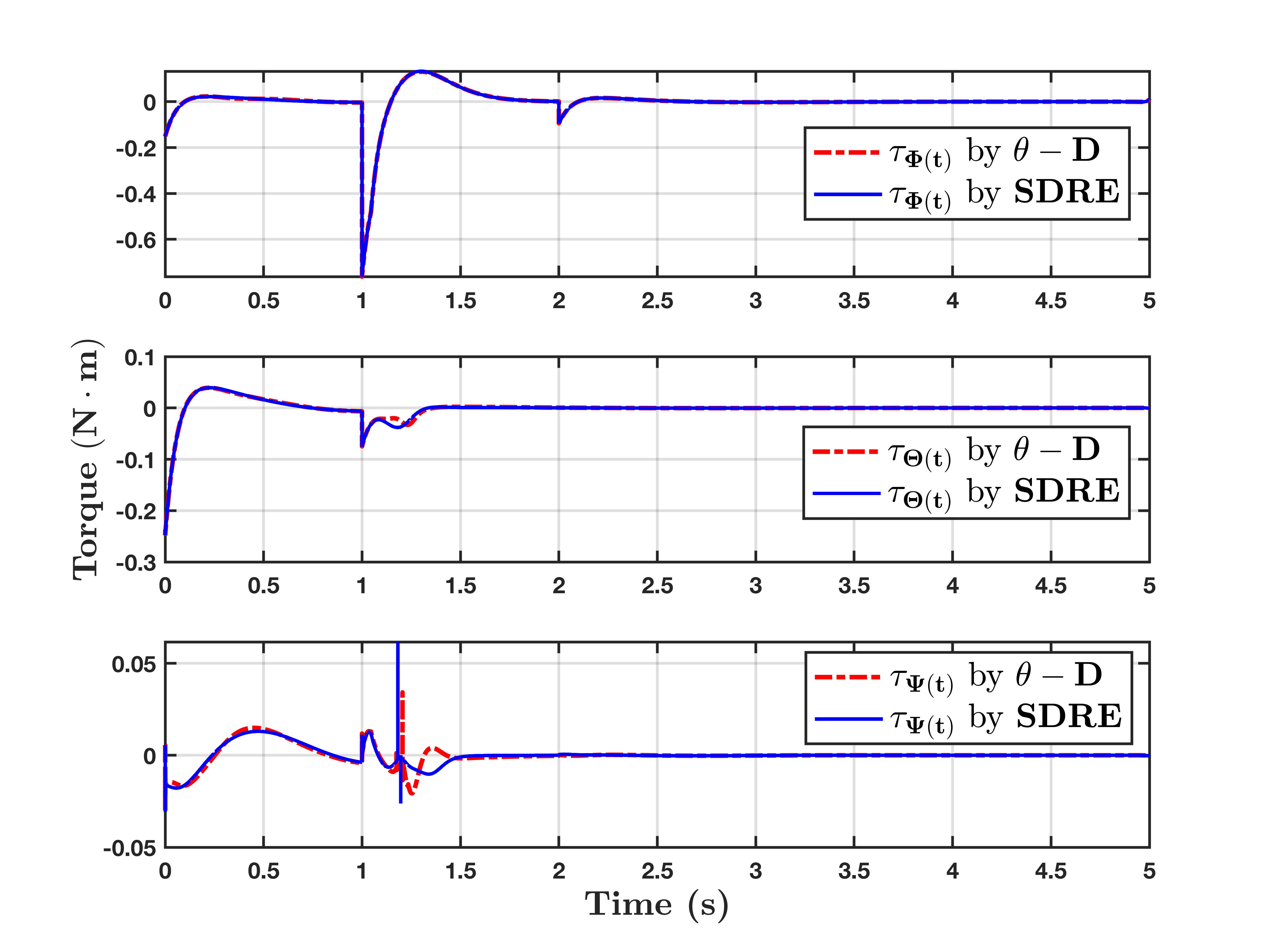}
\caption{ Torque profile by two techniques}
\label{Fig.6}
\end{figure}
\bibliographystyle{IEEEtran}
\bibliography{IEEEabrv,mybibfile}
\begin{figure}
\centering
\includegraphics[scale = 0.057]{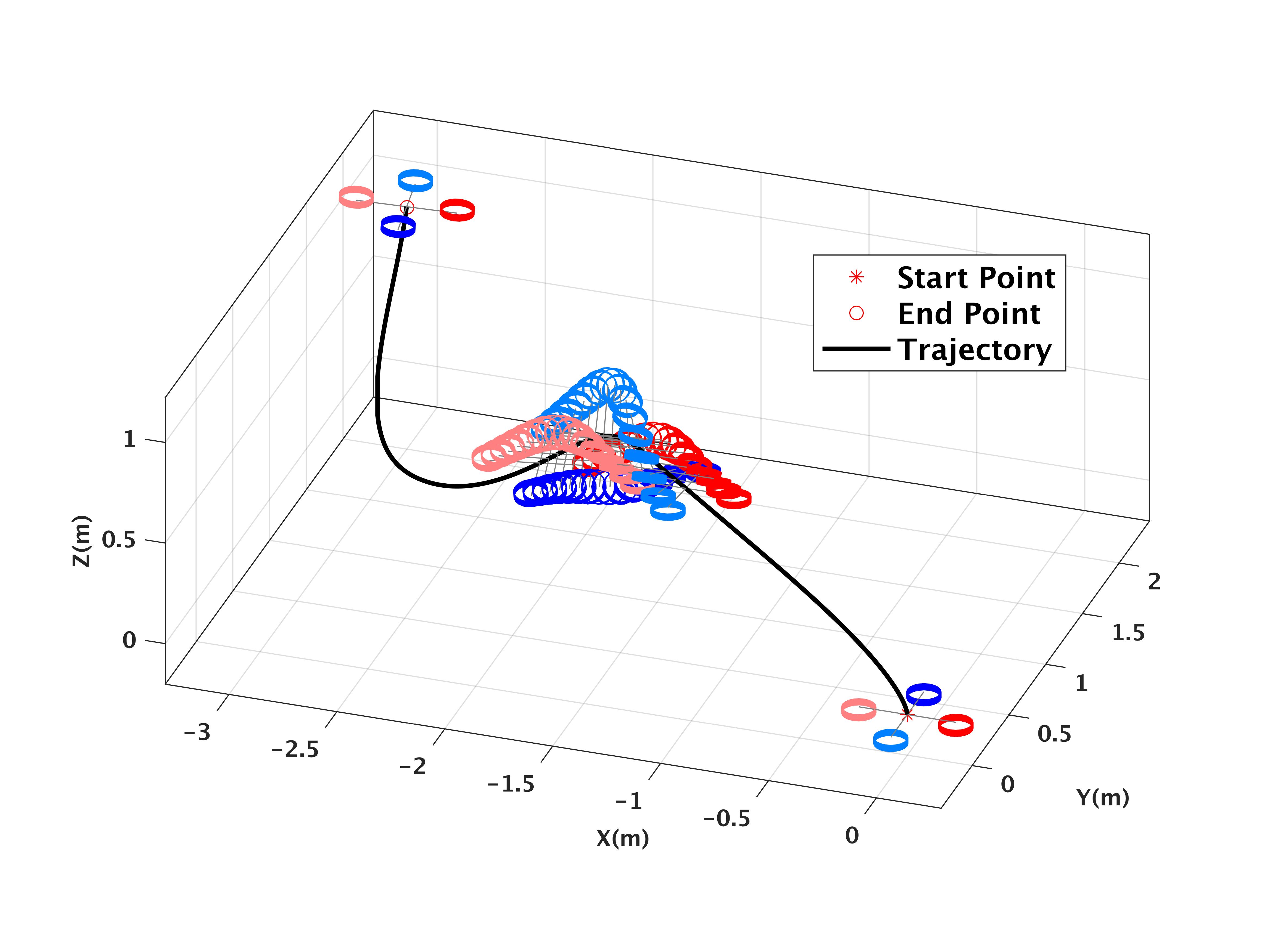}
\caption{Three dimensional motion profile by finite-time $\boldsymbol{\theta}-$D technique}
\label{Fig.7}
\end{figure}
\end{document}